\documentclass[journal]{IEEEtran}
\usepackage{cite}
\usepackage[cmex10]{amsmath}
\usepackage{subfigure}
\ifCLASSINFOpdf
  \usepackage[pdftex]{graphicx}
\else
  \usepackage[dvips]{graphicx}
  \graphicspath{{../figs/}}
  \DeclareGraphicsExtensions{.eps}
\fi

\begin{document}
\title{Demonstration of Spectrum Sensing with Blindly Learned Feature}
\author{Peng~Zhang,~\IEEEmembership{Student Member,~IEEE,}
        Robert~Qiu,~\IEEEmembership{Senior Member,~IEEE,}
        and~Nan~Guo,~\IEEEmembership{Senior~Member,~IEEE}
\thanks{The authors are with the Wireless Networking Systems Lab in Department of Electrical and Computer Engineering, Center for Manufacturing Research, Tennessee Technological University, Cookeville, TN, 38505, USA. E-mail: pzhang21@students.tntech.edu, \{rqiu, nguo\}@tntech.edu}}
\maketitle
\begin{abstract}
Spectrum sensing is essential in cognitive radio. By defining leading \textit{eigenvector} as feature, we introduce a blind feature learning algorithm (FLA) and a feature template matching (FTM) algorithm using learned feature for spectrum sensing. We implement both algorithms on Lyrtech software defined radio platform. Hardware experiment is performed to verify that feature can be learned blindly. We compare FTM with a blind detector in hardware and the results show that the detection performance for FTM is about 3 dB better.
\end{abstract}
\IEEEpeerreviewmaketitle
\section{Introduction}
In cognitive radio, spectrum sensing is required at secondary user (SU) to detect primary user (PU) signal's existence. It needs to work with low signal-to-noise ratio (SNR), noise uncertainty problem \cite{tandra2005fundamental} and unknown channel. In implementation, computation complexity has to be considered as well for real-time performance.

So far, many algorithms have been explored \cite{zeng2010review}, and implemented on hardware as well \cite{chen2010,oh2008tv}. If signal is random and non-white wide-sense stationary (WSS), estimator-correlator (EC), derived from likelihood ratio test (LRT), can be considered as an upper benchmark when signal covariance matrix and noise variance are perfectly known \cite{lim2008glrt}. However, in practice, all parameters are unknown. Maximum to minimum eigenvalue (MME) detector \cite{zeng2007maximum} and covariance absolute value (CAV) algorithm \cite{oh2008tv} are derived and can be considered as a lower benchmark when all parameters are unknown. Moreover, CAV is implemented as well \cite{oh2008tv}. As will be shown in simulation, MME and CAV have similar performance and EC is about 4 dB better than MME/CAV. 

With partial prior knowledge, detection performance should be bounded by EC and MME/CAV. 
We propose to use blindly learned signal feature as prior knowledge so that feature can be local and not restricted to specific signal types. From pattern recognition, the \textit{eigenvectors} 
are considered as features \cite{fukunaga1990introduction}. We define the leading \textit{eigenvector} as signal feature because for non-white WSS signal it is most robust against noise and stable over time. Feature learning algorithm (FLA) and feature template matching (FTM) detector using the blindly learned feature as prior knowledge are developed. We have implemented FLA, FTM and CAV in Lyrtech software-defined-radio platform. Hardware experiment in non-line-of-sight (NLOS) indoor environment shows that feature can be learned blindly. Detection performance of FTM is compared with CAV in hardware experiment, showing $3$ dB improvement. To the best of our knowledge, this is the first \textit{eigenvector} based spectrum sensing approach, as well as hardware implementation.
\section{Problem Statement and Detection Algorithms}
\subsection{Problem Statement}
Spectrum sensing can be modeled as random signal detection problem. Let $r\left( t \right)$ be the continuous-time signal at receiver within channel coherence time, and channel is unknown but static. $r\left( t \right)$ is sampled with period $T_s$, resulting in $r\left[ n \right] = r\left( {nT_s } \right)$. There are two hypothesis:
\begin{equation}
\begin{array}{l}
	\label{Hyp}
 {\bf H}_0 :r\left[ n \right] = w\left[ n \right] \\
 {\bf H}_1 :r\left[ n \right] = s\left[ n \right] + w\left[ n \right]
 \end{array}
\end{equation}
$s\left[ n \right]$ is received PU signal after unknown channel and is zero-mean non-white WSS. $w\left[ n \right]$ is zero-mean white Gaussian noise. Two probabilities are of interest: Detection probability, $P_d \left( {{\bf H}_1 |r \left[ n \right] = s\left[ n \right] + w\left[ n \right]} \right)$, and false alarm probability, $P_f \left( {{\bf H}_1 |r \left[ n \right] = w\left[ n \right]} \right)$.


Let ${\bf r}_n$ be vector consisting of $N$ samples of $r\left[ n \right]$:
\begin{equation}
\label{vector_r}
{\bf r}_n  = \left[ {r\left[ n \right],r\left[ {n + 1} \right], \cdots ,r\left[ {n + N - 1} \right]} \right]^T
\end{equation}
$\left(  \cdot  \right)^T $ denotes matrix transpose. ${\bf r}_n \sim N(0, {\bf R}_r)$, and ${\bf R}_r$ can be approximated by sample covariance matrix ${\hat{\bf R}}_r$. Given the $i^{th}$ sensing segment ${\Gamma _{r,i}} = \left\{ {{{\bf{r}}_i},{{\bf{r}}_{i + 1}}, \cdots {{\bf{r}}_{i + {N_s} - 1}}} \right\}$, we have 
\begin{equation}
\hat{{\bf R}}_{r} = \frac{1}{{N_s }}\sum\limits_{i = 1}^{N_s } {{\bf r}_i {\bf r}_i ^T }
\end{equation}
We will use ${\bf R}_r$ instead of $\hat{{\bf R}}_r$ for convenience. The eigen-decomposition of ${\bf R} _r$ is:
\begin{equation}
\label{eigen_decomp}
\begin{array}{l}
 {\bf R}_r  = \Phi _r \Lambda _r \Phi _r ^T  \\ 
 \Phi _r  = \left[ {\begin{array}{*{20}c}
   {\phi _{r,1} } & {\phi _{r,2} } &  \cdots  & {\phi _{r,N} }  \\
\end{array}} \right] \\ 
 \Lambda _r  = diag\left\{ {\lambda _{r,1} ,\lambda _{r,2} , \cdots ,\lambda _{r,N} } \right\} \\ 
 \end{array}
\end{equation}
where $diag\left\{ \cdot \right\}$ denotes the diagonal matrix, $\left\{ {\phi _{r,i} } \right\}$ are eigenvectors of ${\bf R}_r$ and $\left\{ {\lambda _{r,i} } \right\}$ are eigenvalues of ${\bf R}_r$, satisfying $\lambda _{r,1}  \ge \lambda _{r,2}  \ge ... \ge \lambda _{r,N}$. Accordingly, we have signal covariance matrix ${\bf R}_s$, noise covariance matrix ${\bf R}_w = \sigma^2 {\bf I}$, where ${\bf I}$ is identity matrix.
\subsection{Detectors with All Parameters Known or Unknown}
Assume the $i^{th}$ sensing segment $\Gamma_{r,i}$ is available. If parameters ${\bf R}_s$ and $\sigma^2$ are perfectly known, we have the optimum EC detector for ${\bf r}_i$, derived from LRT \cite{lim2008glrt}. ${\bf H}_1$ is true if 
\begin{equation}
{T_{EC}}\left( {{{\bf{r}}_i}} \right) = {\bf{r}}_i^T{{\bf{R}}_s}{\left( {{{\bf{R}}_s} + {\sigma ^2}{\bf{I}}} \right)^{ - 1}}{{\bf{r}}_i}  > \gamma
\end{equation}
where $\gamma$ is the threshold determined by desired $P_f$ ($\gamma$ has same meaning in the rest of this letter). We perform EC on all vectors in $\Gamma_{r,i}$ and then do average. Take $\Gamma_{r,1}$ for example. ${\bf H}_1$ is true if
\begin{equation}
{{\bar T}_{EC}}\left( {{\Gamma _{r,1}}} \right) = \frac{1}{{{N_s}}}\sum\limits_{i = 1}^{{N_s}} {{\bf{r}}_i^T{{\bf{R}}_s}{{\left( {{{\bf{R}}_s} + {\sigma ^2}{\bf{I}}} \right)}^{ - 1}}{{\bf{r}}_i}} > \gamma
\end{equation}
It is impractical to assume ${\bf R}_s$ and $\sigma^2$ known due to the unknown channel and noise uncertainty problem. In \cite{zeng2007maximum}, MME is derived, assuming all parameters unknown. Note that there is another MME derived from General LRT (GLRT) \cite{lim2008glrt}. Though they have the same formula, \cite{lim2008glrt} is for multiple receive antennas, while \cite{zeng2007maximum} is for single receive antenna. Here MME refers to \cite{zeng2007maximum}. MME gets ${\bf R}_r$ from $\Gamma_{r, i}$ and ${\bf H}_1$ is true if 
\begin{equation}
{T_{MME}}\left( {{\Gamma _{r,i}}} \right) = \frac{{{\lambda _{r,1}}}}{{{\lambda _{r,N}}}} > \gamma
\label{T_MME}
\end{equation}
MME is totally blind and does not have noise uncertainty problem. However, MME needs to calculate maximum and minimum eigenvalues, which is not implementation friendly. An alternative detector, CAV, has been proposed and implemented \cite{oh2008tv}. As will be shown in simulation, CAV has almost the same performance with MME. CAV gets ${\bf R}_r$ from $\Gamma_{r, i}$ and calculates two parameters: 
\[{T_1} = \frac{1}{N}\sum\limits_{i = 1}^N {\sum\limits_{j = 1}^N {\left| {{r_{ij}}} \right|} } ;
{T_2} = \frac{1}{N}\sum\limits_{i = 1}^N {\left| {{r_{ii}}} \right|} \]
where $r_{ij}$ are elements of ${\bf R}_r$. Then, ${\bf H}_1$ is true if
\begin{equation}
\label{T_CAV}
{T_{CAV}}\left( {{\Gamma _{r,i}}} \right) = \frac{{{T_1}}}{{{T_2}}} > \gamma
\end{equation}
\subsection{Feature Learning based Spectrum Sensing}
With partial prior knowledge, detection performance should be bounded by EC and MME/CAV. We introduce blind feature learning and use feature as prior knowledge for detection. In pattern recognition \cite{fukunaga1990introduction}, eigenvectors are called signal features and the leading eigenvector has greatest mutual information with original signal. We assume ${\bf R}_s$ to be rank-1 matrix and define feature $\varphi_s$ as the leading eigenvector $\phi_{s,1}$ only. It is well known in pattern recognition that feature is random for white noise while stable for non-white WSS signal. Moreover, feature is most robust against noise \cite{fukunaga1990introduction}. Therefore if PU signal exists, highly similar features can be detected in consecutive sensing segments $\Gamma_{r, i}$. This phenomenon will be shown in simulation with real-world data. Due to the robustness of signal feature, we can learn it blindly. A blind feature learning experiment in NLOS environment will be demonstrated.

We develop FLA for blind feature learning based on $2$ consecutive sensing segments $\Gamma _{r,i}$ and $\Gamma_{r,i+1}$. 
\begin{enumerate}
	\item Compute corresponding features $\varphi_i$ and $\varphi_{i+1}$
	\item Compute feature similarity $\rho$ via template matching:
		\begin{equation}
		\label{similarity}
		\rho_{i, i+1} = \mathop {\max }\limits_{l = 1,2,...,N - k + 1} | {\sum\limits_{k = 1}^N {\varphi _i \left[ k \right]\varphi _{i+1} \left[ {k + l} \right]} } |
		\end{equation}
	\item If $\rho_{i, i+1} > T_e$, signal feature $\varphi_s$ is learned as $\varphi_{i+1}$, where $T_e$ is determined by $\rho_{i,i+1}$ of pure noise.
\end{enumerate}
Because noise features are random while signal features are stable, FLA can learn $\varphi_s$ accurately with high $T_e$, which is independent of signal energy and/or noise energy. When $\varphi_s$ is learned, we have the FTM detector:
\begin{enumerate}
	\item Extract feature $\varphi _{current}$ from $\Gamma_{r, current}$.
	\item Compute $\rho_{current, s}$ of $\varphi _{current}$ and $\varphi_s$.
	\item ${\bf H}_1$ is true if
	\begin{equation}
{T_{FTM}}\left( {{\Gamma _{r,i}}} \right) = {\rho _{current,s}} > \gamma
	\label{T_FTM}
	\end{equation}
where $\gamma$ is the threshold determined by desired $P_f$.
\end{enumerate}
\section{Simulation Results}
We first demonstrate that signal feature is stable while noise feature is random. Field measurements of DTV done in Washington D.C. \cite{DTV2006Measurements} are used as PU signal for illustration purpose. The captured signal has a duration of about 25 seconds. Receiver SNR and the communication channel between the transmitter and receiver are unknown. We divide the samples into $5000$ sensing segments and use FLA to calculate consecutive feature similarities for both signal and noise separately. In the simulation, $N_s = 10^5$ and $N = 32$. By setting $T_e = 90\%$, $\rho_{i, i+1} > T_e$ for $99.46\%$ in all segments when PU signal exists. Moreover, the similarity between the features of the first sensing segment and the last sensing segment is $99.98\%$, showing that signal feature is very stable and almost unchanged in 25 seconds. If PU signal does not exist, $\rho_{i, i+1} > T_e$ for only $0.82\%$, meaning that noise features are random. Currently $T_e$ is set empirically. More efforts will be made to determine $T_e$ analytically.

\begin{figure}[tbp]
	\centering
		\includegraphics[width=0.45\textwidth]{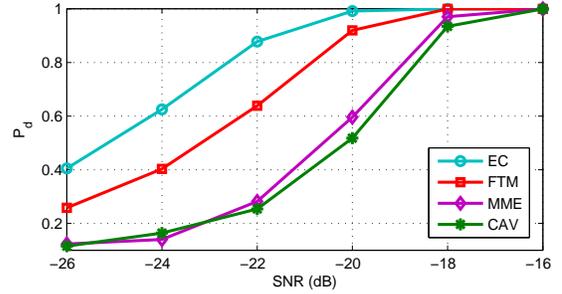}
	\caption{$P_d$ vs. SNR at $P_f = 10\%$ for various detection algorithms.}
	\label{fig:Detection_DTV_SNR_All}
\end{figure}

%
In the $2^{nd}$ simulation, we pick the last signal feature as $\varphi_s$. One segment of the captured DTV data is used to simulate the detection performance of EC, FTM, MME and CAV with $N_s = 10^5$ and $N = 32$. Captured data are considered as clean signal and noise is added according to different SNR level. $1000$ simulations are performed for each SNR plot. Fig. \ref{fig:Detection_DTV_SNR_All} shows $P_d$ vs. SNR at $P_f = 10\%$. MME and CAV have almost the same performance and FTM is in between of two benchmarks. To reach $P_d \approx 100\%$, minimum required SNR for LRT is about $-20$ dB, FTM is about $-18$ dB, MME/CAV is about $-16$ dB.
\section{Hardware Implementation and Experiment}
\begin{figure}[tbp]
	\centering
 		\includegraphics[width=0.45\textwidth]{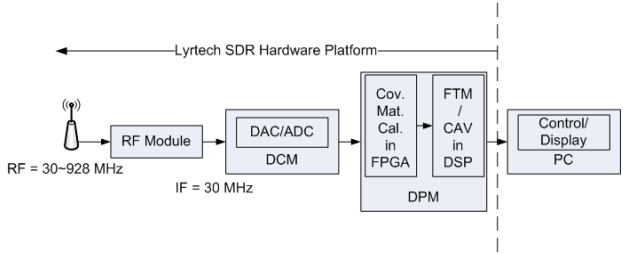}
	\caption{The top-level architecture of the spectrum sensing receiver.}
	\label{fig:Arch}
\end{figure}

We implement FLA, FTM and CAV in the Lyrtech platform. The platform has three modules: Tunable RF module, data conversion module and digital processing module. The algorithms are implemented in digital processing module with Xilinx FPGA and TI DSP. The top-level architecture of the implementation is shown in Fig. \ref{fig:Arch}. 
Covariance matrix calculation is implemented in FPGA, shared by all algorithms. Parameters for the calculation are $N_s = 2^{20}$ and $N = 32$. Leading eigenvector calculation, $T_1$ and $T_2$ calculation are implemented in DSP. Leading eigenvector calculation involves eigen-decomposition. To achieve low computation complexity, we use the Fast-PCA (FPCA)\cite{sharma2007fast}, whose computation complexity is ${\cal O}(N^2)$. This is especially low for $N = 32$. Without any effort in code optimization, leading eigenvector calculation can be done within $20$ ms. 

We first demonstrate blind feature learning in NLOS environment. We use Rohde \& Schwarz signal generator as the PU transmitter and Lyrtech platform as the SU receiver. Transmit antenna and receive antenna are 2 meters away, and the direct path is blocked by the signal generator. A $-50$ dBm sinusoidal signal at $435$ MHz is transmitted to emulate PU signal. SU's RF is tuned to $432$ MHz center frequency with $20$ MHz bandwidth. We turn on the signal generator and use FLA to measure similarities for $20$ seconds in hardware. Receiver is totally blind and the environment is interference-free. 
By setting $T_e = 80\%$, $\rho_{i, i+1} > T_e$ for $87.6\%$ amount of time. Moreover, the similarity between the features of the first sensing segment and the last sensing segment is $94.3\%$, indicating that PU feature is almost unchanged after $20$ seconds. As a result, PU feature in this experiment is very stable and can be learned blindly. We pick the feature of the last sensing segment as $\varphi_s$ for the next experiment.

\begin{figure}[tbp]
	\centering
		\includegraphics[width=0.45\textwidth]{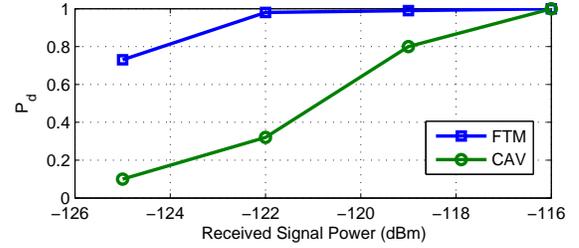}
		\caption{$P_d$ vs. Received Signal Power at $P_f = 10\%$ for FTM and CAV.}
	\label{fig:Detection_Hardware}
\end{figure}

In this experiment we compare the detection performance of FTM and CAV in hardware. FTM has PU feature stored as $\varphi _{s}$ from the previous experiment. In order to compare the $P_d$ of both algorithms under stable received signal power, we connect the signal generator to the receiver with SMA cable. Transmit power is set from $-125$ dBm to $-116$ dBm with $3$ dB increments. Cable loss is omitted and received signal power is measured at the transmitter. $1000$ tests are made in hardware for each setting. Fig. \ref{fig:Detection_Hardware} shows the $P_d$ vs. received signal power curves at $P_f = 10\%$. It can be seen that FTM requires at least $-119$ dBm to reach $P_d = 100\%$, while CAV requires at least $-116$ dBm, $3$ dB higher than FTM.
\section{Conclusion}

To the best of our knowledge, for the first time, the leading \textit{eigenvector} of signal is used as feature to improve detection performance. FLA is proposed for blind local feature learning and FTM is proposed for signal detection, using the blindly learned local feature. The detection performance of FTM lies within EC, the upper benchmark when all parameters known, and MME/CAV, the lower benchmark when all parameters unknown. We have implemented FLA, FTM and CAV in Lyrtech software-defined-radio platform. We use simulation and hardware experiment to verify that feature can be learned blindly. We compare detection algorithms in simulation and hardware as well. With learned feature as prior knowledge, FTM is about $3$ dB better than the blind detection algorithm. 

This is the first step in the `intelligent' CR receiver design. More work will be done in fast and robust learning and spectrum sensing with multiple PU/interference. Computation efficient methods for large size sample covariance matrix \cite{frieze2004fast} will also be explored. 
\section*{Acknowledgment}
This work is funded by National Science Foundation through grants (ECCS-0901420), (ECCS-0821658), and Office of Naval Research through two contracts (N00014-07-1-0529, N00014-11-1-0006).

\bibliographystyle{ieeetr}
\bibliography{bib/CR,bib/Pattern_Recognition} 

\ifCLASSOPTIONcaptionsoff
  \newpage
\fi

\end{document}